\documentclass[aps, prd, twocolumn, superscriptaddress, nofootinbib]{revtex4}
\usepackage{graphicx}% Include figure files
\usepackage{dcolumn}% Align table columns on decimal point
\usepackage{amssymb}% outlined letters

\begin{document}

%Macros
\newcommand{\Eq}[1]{\mbox{Eq. (\ref{eqn:#1})}}
\newcommand{\Fig}[1]{\mbox{Fig. \ref{fig:#1}}}
\newcommand{\Sec}[1]{\mbox{Sec. \ref{sec:#1}}}

\newcommand{\PHI}{\phi}
\newcommand{\PhiN}{\Phi^{\mathrm{N}}}
\newcommand{\vect}[1]{\mathbf{#1}}
\newcommand{\Del}{\nabla}
\newcommand{\unit}[1]{\;\mathrm{#1}}
\newcommand{\x}{\vect{x}}
\newcommand{\ScS}{\scriptstyle}
\newcommand{\ScScS}{\scriptscriptstyle}
\newcommand{\xplus}[1]{\vect{x}\!\ScScS{+}\!\ScS\vect{#1}}
\newcommand{\xminus}[1]{\vect{x}\!\ScScS{-}\!\ScS\vect{#1}}
\newcommand{\diff}{\mathrm{d}}

\newcommand{\be}{\begin{equation}}
\newcommand{\ee}{\end{equation}}
\newcommand{\bea}{\begin{eqnarray}}
\newcommand{\eea}{\end{eqnarray}}
\newcommand{\vu}{{\mathbf u}}
\newcommand{\ve}{{\mathbf e}}

        \newcommand{\vU}{{\mathbf U}}
        \newcommand{\vN}{{\mathbf N}}
        \newcommand{\vB}{{\mathbf B}}
        \newcommand{\vF}{{\mathbf F}}
        \newcommand{\vD}{{\mathbf D}}
        \newcommand{\vg}{{\mathbf g}}
        \newcommand{\va}{{\mathbf a}}

%=====================================================================
%=====================================================================
%=====================================================================

\title{Saddle stresses for generic theories with a preferred acceleration 
scale}
%Opening up the parameter space of MOND}

\newcommand{\addressImperial}{Theoretical Physics, Blackett Laboratory, Imperial College, London, SW7 2BZ, United Kingdom}

\author{Jo\~{a}o Magueijo}
\email{magueijo@ic.ac.uk}
\affiliation{\addressImperial}

\author{Ali Mozaffari}
\email{ali.mozaffari05@ic.ac.uk}
\affiliation{\addressImperial}

\date{\today}

\begin{abstract}
We show how scaling arguments may be used to generate 
templates for the tidal stresses around saddles for a vast class of
MONDian theories {\it detached from their obligations as dark matter 
alternatives}. Such theories are to be seen simply as alternative theories 
of gravity with a preferred acceleration scale, 
and could be tested in the solar system by extending the LISA 
Pathfinder (LPF) mission. The constraints thus obtained may then be combined,
if one wishes, with requirements arising from astrophysical and cosmological 
applications, but a clear separation of the issues is achieved.
The central technical content of this paper is the derivation of 
a scaling prescription allowing complex numerical work
to be bypassed in the generation of templates. We find that LPF could 
constrain very tightly the acceleration $a_0$ and the free parameter $\kappa$
present in these theories.
As an application of our technique we also produce predictions for the 
moon saddle (for which a similar scaling argument is applicable) with
the result that we recommend that it should be included in orbit design.
\end{abstract}


\maketitle

%=====================================================================
%=====================================================================
%=====================================================================

\section{Introduction}
There is an ubiquitous acceleration scale in the universe, 
$ a_0\sim 10^{-10}\unit{ms}^{-2}$, which
turns up variously in cosmology and astrophysics: the  
cosmic expansion rate, galactic rotation curves, etc. This observation
has prompted the investigation of alternative theories of gravity 
endowed with a preferred acceleration.
TeVeS~\cite{teves}, and more generally 
relativistic MONDian 
theories~\cite{BSTV,aether,aether1,Milgrom:2009gv,Milgrom:2010cd}, 
provide a blueprint for such 
constructions. MONDian theories were first proposed with the motivation
of bypassing the need for dark matter~\cite{Milgrom:1983ca,Fam-gaugh}.
However, they may also
be considered independently from this application, and be seen simply
as alternative theories of gravity~\cite{pedroreview}
into which an acceleration scale has been embedded. 
In this guise they constitute 
prime targets for experimental gravitational tests inside the solar system.

As an example let us consider TeVeS (but what follows applies 
generally to what in~\cite{ali} was labelled ``Type I''
theories). Abstracting from aspects which do not affect the 
non-relativistic limit, 
the theory benefits from the leeway of a whole free-function
$\mu$. Its choice may be informed by minimalism and simplicity,
e.g. $\mu$ may be built to encode only 2, rather than 3 or
more regimes. Putting aside details affecting the transition
between the two regimes, we are then left with two free parameters: 
$a_0$ (the acceleration scale of the theory) and $\kappa$ (controlling 
the renormalization of the gravitational constant $G$). These are 
fixed by astrophysical and cosmological applications, if 
the theory is to act as a competitor to dark matter. 
But $a_0$ and $\kappa$ can also be seen as fully free parameters 
in any solar system test. 
%One could then separate the experimental
%pressures upon the theory, distinguishing direct tests from the more 
%uncertain and model dependent constraints arising from galactic and 
%extra-galactic observations. 

Specifically, these theories predict a rich phenomenology around the saddle
points of the gravitational potential. The prospect of extending the 
LISA Pathfinder mission so that a saddle of the Sun-Earth-Moon system
is visited brings them within experimental striking 
range~\cite{ali,bevis,bekmag,companion,aliqumond,TypeII}. 
Predictions for minimal theories constrained by 
cosmological and astrophysical applications were studied in~\cite{ali},
where the general impact of a negative result 
was also examined. Detaching the target theory from its duties as 
``dark matter'' alternative requires the generation of a large database of
templates. However, re-running the adaptive-mesh code  presented 
in~\cite{bevis} for each $\mu$ is simply not feasible, 
and we run up against a computational wall.

In this paper we show how this work can be partly alleviated. 
As long as we are interested in changing only $\kappa$ and $a_0$, 
a simple scaling argument allows the
generation of the whole set of required templates from those obtained
with fiducial values for $a_0$ and $\kappa$, shortcutting tedious or 
downright impossible hard labour. The analytical argument is laid down in
Section~\ref{scale}, and its application to LPF is given 
in Section~\ref{application}. In order to include another topical
application, in Section~\ref{moon} we also show how the lunar saddle
would fare, were LPF to include it in a mission extension. 

\section{Scaling behaviour around saddles}\label{scale}
Scaling is an interesting tool for generating solutions to apparently
intractable problems. For example imposing a self-similar ansatz leads to 
striking progress in the study of gravitational collapse, rendering what {\it a priori} are PDEs into simpler ODEs (e.g.~\cite{BertSelfSim,BertSelfSimSec}).
%[find the bertschinger homothetic solutions - he calls then self-similar
%and it's r/t; check it's ApJ *58 pp 39 (1985), etc].
Scaling 
behaviour was observed in the MONDian tidal stresses around saddles, when
comparing the profiles around the Moon saddle and the Earth-Sun saddle
(see Fig. 12 in~\cite{bevis}, and its surrounding comments). It was noted that
the tidal stresses are very approximately the same
once they are spatially stretched and their amplitude scaled 
to account for the different Newtonian tidal stress $A$. 
In what follows we rigorously explain this empirical fact and extend
its scope, deriving the scaling laws associated with varying 
$a_0$ and $\kappa$.

In TeVeS and other type I theories~\cite{ali} the non-relativistic dynamics
results from the joint action of the usual Newtonian potential
$\Phi_N$ (associated with the metric) and a ``fifth force'' 
scalar field, $\phi$, responsible
for MONDian effects. The total potential acting on
non-relativistic particles is their sum $\Phi=\Phi_N+\phi$.
The field $\phi$ is ruled by a non-linear Poisson equation: \be \nabla \cdot
\left(\mu(z)\nabla \phi\right) = \kappa G \rho, \ee with
$z=\frac{\kappa}{4\pi}\frac{\vert\nabla\phi\vert}{a_0}$. 
Here $\kappa$ is a dimensionless constant and $a_0$ is the usual MONDian
acceleration. 

For these theories, the argument presented in Sections II and 
IV of~\cite{bekmag}, for a specific $\mu$, can 
be generalized for any $\mu$. It is always possible to define a variable:
\be\label{Udef}
\mathbf{U}=-\frac{\kappa}{4\pi a_0}\mu(z)\nabla\phi
\ee
in terms of which the vacuum equations become:
\bea
\nabla\cdot \mathbf{U}=0
\label{a}
\\
\alpha(U)\,U^2\,\nabla\wedge \mathbf{ U}+\mathbf{ U}\wedge\nabla
U^2=0, \label{b} \eea
i.e. the free parameters $a_0$ and $\kappa$ drop out, and
``universal'' equations are obtained. Here
\be
\alpha(U)=\frac{d\ln U^2}{d\ln \mu}\; ,
\ee
(notice that $U^2=z^2\mu^2(z)$, so that $\mu$ can be written as 
a function of $U$ alone, $\mu=\mu(U)$). The MONDian physical force can be
obtained from $\mathbf{U}$ using:
\be
\mathbf{F}_\phi= -\nabla\phi
={4\pi a_0\over k}\frac{1}{\mu(U)}\mathbf{U}\; ,
\ee
which follows directly from (\ref{Udef}).
In~\cite{bekmag} the choice was made:
\be
\mu=\frac{U^{1/2}}{(1+U^2)^{1/4}}
\ee
so that $\alpha=4(1+U^2)$,
in agreement with Eqns.~(30)-(31) of~\cite{bekmag}, and
\be   \mathbf{F}_\phi=           -\nabla\phi
={4\pi a_0\over k}(1+U^2)^{1/4}{\mathbf{U}\over
U^{1/2}}\,, \label{gradphi} \ee
in agreement with Eqn.~32 in~\cite{bekmag}. However, as we see, the argument 
can be adapted to any function $\mu$.

Equations (\ref{a}) and (\ref{b}) are invariant under
a rigid rescaling of the spatial variables:
\bea
\mathbf{U}&\rightarrow& \mathbf{U}\\
\mathbf{x}&\rightarrow& \lambda \mathbf{x}
\eea
where $\lambda$ is spatially constant. To use the technical term
they admit homothetic solutions, i.e.:
\be\label{homot}
\mathbf{U}=\mathbf{F}(\lambda \mathbf{x})\; .
\ee
where $\mathbf{F}$ is a universal function. However, 
we have yet to supply Equations (\ref{a}) and (\ref{b}) with boundary
conditions. This is done by going far enough from the saddle so
that the field $\phi$ has entered the Newtonian regime. 
With the conventions used in~\cite{ali} one has $\mu\rightarrow 1$
(the renormalization in $G$ is fully absorbed in $\kappa$), and so:
\be
\phi\approx \frac{\kappa}{4\pi}\Phi_N\; .
\ee
The appropriate boundary condition is then supplied from the 
Newtonian limit relation:
\be
\mathbf{U}\approx \frac{\kappa}{4\pi a_0}\mathbf{F}_\phi\approx
{\left(\frac{\kappa}{4\pi}\right)}^2\frac{1}{a_0}\mathbf{F}_N\; .
%\approx
%{\left(\frac{\kappa}{4\pi}\right)}^2\frac{Ar}{a_0}\mathbf{N}\; .
\ee
Let us first assume that we can approximate the Newtonian
field around the saddle as a linear function, for the 
purpose of effectuating this matching. Then:
% (although this
%simplification will be dropped later). Specifically, let:
\be\label{newtonian1} 
{\mathbf F}_N=-\nabla \Phi_N=Ar{\mathbf N}(\theta,\phi), 
\ee
where $A$ is the Newtonian tidal stress at the saddle point,
and  $\mathbf{N}$  is its angular profile (see Eqs.~35-37
in~\cite{bekmag}). Defining the  MONDian ``bubble size'' as usual:
\be
r_0=\frac{16 \pi ^2 a_0}{\kappa^2 A} \label{bubble}
\ee
we therefore have in the Newtonian regime and close enough to the
saddle:
\be
\mathbf{U}\approx \frac{r}{r_0}\mathbf{N}(\theta,\phi)\; .\label{U Newt}
\ee
This boundary condition allows us to select the homothetic
solution (\ref{homot}) appropriate to a given saddle and free parameters. 
To match the boundary
conditions one should set $\lambda=1/r_0$, so that the solution is
\be
\mathbf{U}=\mathbf{F}{\left(\frac{\mathbf x}{r_0}\right)}\; .
\ee
The above argument is still (approximately) valid if one 
goes beyond the linear approximation, as long as this approximation is
good up to a few $r_0$. If the parameters $a_0$ and $\kappa$ lead to
a breakdown  of this assumption, however, then scaling is lost.

We can now read off similar scaling laws for more familiar 
quantities. Using (\ref{gradphi}) we see that the MONDian
force must have the form:
\be
\mathbf {F}_\phi=
\frac{a_0}{\kappa} {\mathbf G}{\left(\frac{\mathbf x}{r_0}\right)}
\ee
where $\mathbf G$ is another universal function. (This scaling law is obvious
by direct inspection of the analytical solutions derived for the $\mu$ 
used in~\cite{bekmag}; however, as we now see, it is more general).
By taking derivatives 
we then find that the MONDian tidal stresses must have the form
\be
S_{ij}=\kappa A H_{ij}{\left(\frac{\mathbf x}{r_0}\right)} \label{ts}
\ee
where the  $H_{ij}$ are also universal. This explains the 
scaling law observed for different $A$, and fixed $\kappa$ and $a_0$,
when comparing the Moon and Earth saddles~\cite{bevis}.
But it also allows for templates for general values of $\kappa$ and $a_0$ 
to be generated from those for fiducial values simply by rescaling 
them according to the above laws.

%[TO BE DONE]
%A similar argument can be applied to the Newtonian forces, 
%which also take homothetic solutions. The argument can then
%be made rigorously, beyond the linear approximation. 
%I suspect that the expression for $r_0$
%may change slightly, but the scaling argument is indeed
%rigorous. In fact U may change too. This scaling is lost as explained above.

\section{An application}\label{application}
The practical applications of the previous section are far-reaching
and will be the subject of a number of future publications devoted to
the data analysis of a saddle test. 
As a simple example we examine in this section the impact of 
$a_0$ and $\kappa$ on the SNR (Signal to Noise Ratio) forecast for a LISA
Pathfinder flyby. As in~\cite{ali}, we assume the use of an optimal
noise-matched filter, using for our noise model the best  
estimate at the time of writing (labelled 
``Best Case Noise'' in Fig.6 of~\cite{ali}). We then inspect the SNR variations with $a_0$ and $\kappa$ for
different saddle impact parameters $b$. After a number of studies, following 
on from~\cite{companion}, an impact parameter $b\sim 10-50 \unit{km}$ is now considered
realistic. Multiple flybys are currently being investigated, for which
$b$ may not be as good. We therefore consider SNRs for
$b$ up to 1000~km. Recall that for the fiducial 
values $a_0=10^{-10}\unit{ms}^{-2}$ and $\kappa=0.03$ (required,
or suggested, by cosmological and
astrophysical applications) one forecasts SNRs for the Earth-Sun saddle
around 40-60 for the expected $b=10-50\unit{km}$, 
only dropping below 5 beyond $b\sim 700\unit{km}$
(see Fig.7 of~\cite{ali}).

%Also when would we lose the ability to detect the theory, should the parameters
%be different from the canonical values suggested by cosmological applications.
%\begin{figure}\begin{center}
%\resizebox{1.\columnwidth}{!}{\includegraphics{bestworstrevised.eps}}
%\caption{\label{fig:noise}{Amplitude Signal Density plot of the noise model used,
%here we picked the best predicted noise model.}}\end{center}
%\end{figure}

\begin{figure}
\begin{center}
\resizebox{1.\columnwidth}{!}{\includegraphics{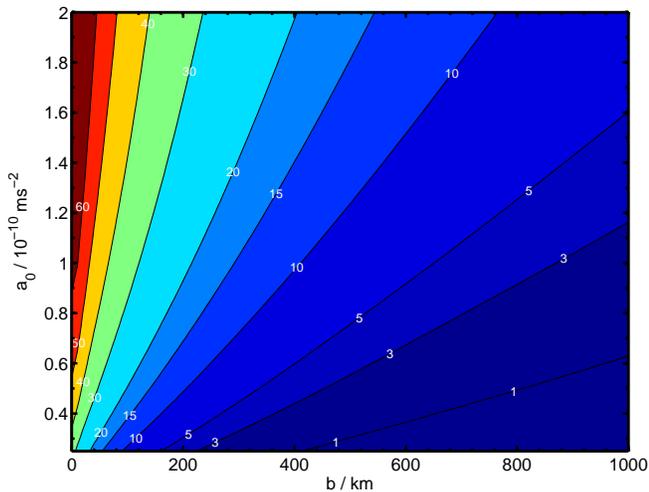}}
\caption{\label{fig:da}{The effect on the Signal-to-Noise Ratio
(SNR) resulting from varying $a_0$,
assuming different impact parameters $b$ and the best estimate for the
noise at the time of writing (with $\kappa$ kept fixed at 
$\kappa=0.03$). The fiducial value used in
previous publications is $a_0 = 10^{-10}\unit{ms}^{-2}$. Generally the 
larger the $a_0$ the higher the SNR.  }}\end{center}
\end{figure}

\begin{figure}\begin{center}
\resizebox{1.\columnwidth}{!}{\includegraphics{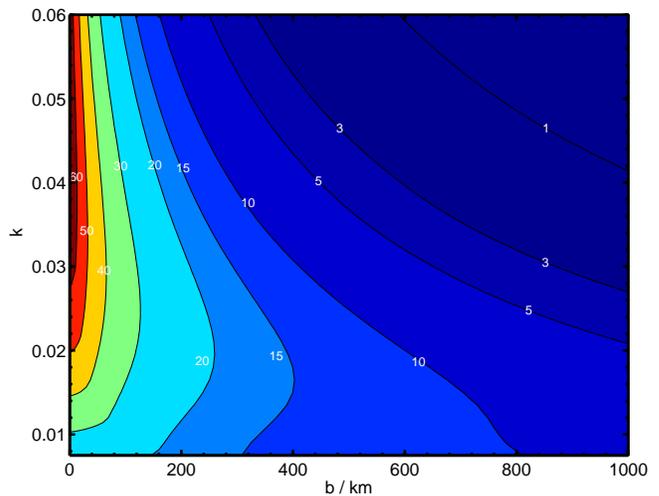}}
\caption{\label{fig:dk}{Effect on the SNR obtained by varying $\kappa$
(keeping $a_0$ fixed at the fiducial value). The 
fiducial value used in previous publications is $\kappa = 0.03$.  
At small $b$, changing $\kappa$ may increase or decrease the SNR (see text
for explanation). At large $b$ one is better off with a small
$\kappa$.}}\end{center}
\end{figure}

\begin{figure}\begin{center}
\resizebox{1.\columnwidth}{!}{\includegraphics{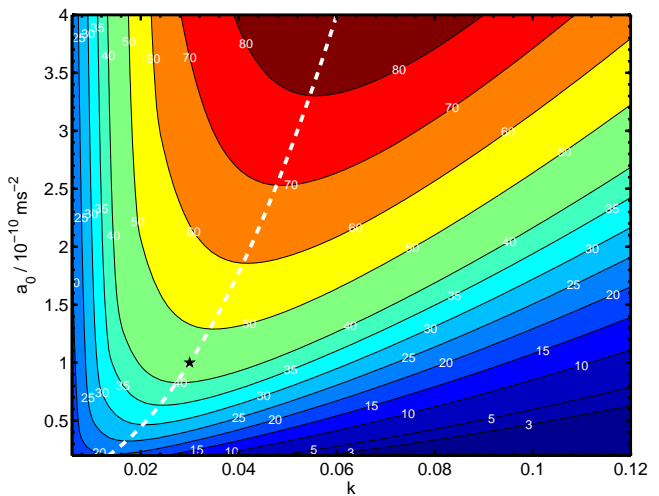}}
\caption{\label{fig:dak}{Effect on SNR obtained by jointly
varying $a_0$ and $\kappa$, for a trajectory with impact parameter
$b=50$km. The fiducial values have been indicated with a star.  
We also plot (dashed white line) the contour of constant $r_0$
passing through the fiducial values.}}\end{center}
\end{figure}

The effect of changing the acceleration scale 
$a_0$ is plotted in Fig.~\ref{fig:da}.
It results from a change in the MOND bubble
size $r_0$, as predicted by Eqn.~\ref{bubble}.
Therefore the SNR is roughly constant 
on lines of constant $b/a_0$. The slope of the iso-SNR lines is not constant
and they are not exactly straight because the SNR algorithm is quite
complicated and non-linear. We see that even at large $b$ it is possible 
to turn a weak result into a strong positive one by increasing $a_0$ by 
a factor of 2. Conversely, if $a_0$ is halved, a SNR below $\sim 5$ is now 
a liability for  $b$ as low as $\sim 350\unit{km}$. Without external 
constraints fixing $a_0$ to better than an order of magnitude, it is
therefore risky to give up on a $b\sim 10-50 \unit{km}$.

The effect of changing $\kappa$ is plotted in Fig.~\ref{fig:dk}
and results from two sources: a change in bubble size according to 
$r_0\propto 1/\kappa^2$ (cf. Eqn.~\ref{bubble}) and an overall factor 
multiplying the amplitude (cf. Eqn.~\ref{ts}). 
The two effects counteract each other, so that unless $b$ is very large, 
the SNR at first increases with $\kappa$, then decreases. For the 
expected $b\sim 10-50$ it can go either way. For large $b$
(greater than $b\sim 500\unit{km}$ for the fiducial value of $a_0$), 
the bubble size prevails and so the SNR
decreases with increasing $\kappa$. The interplay of these two effects is 
best illustrated in Fig.~\ref{fig:dak}, where we plotted the effect 
on the SNR of changing simultaneously $a_0$ and $\kappa$ for 
fixed $b=50\unit{km}$.  We also plotted the line of constant
$r_0$ passing through the fiducial values. As we see the SNR does change
along this line, showing that the bubble size $r_0$ is not the only consideration.
%end of the
%story. 
%We have leverage to either boost or kill off 
%the perspective of a detection by changing the parameters by an
%order of magnitude.  

\begin{figure}\begin{center}
\resizebox{1.\columnwidth}{!}{\includegraphics{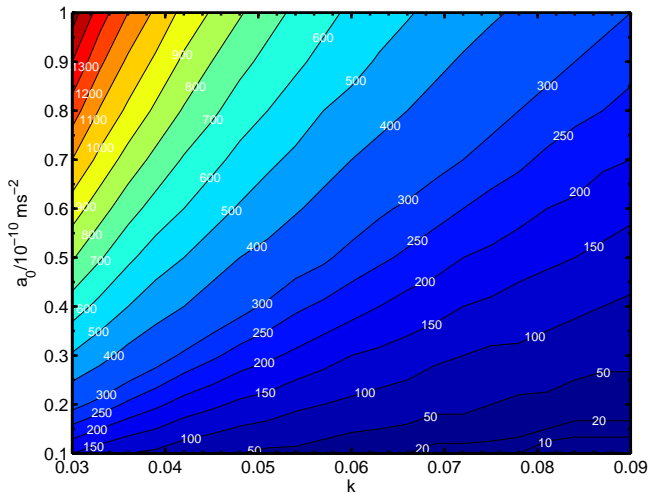}}
\caption{\label{fig:null}{Constraints placed on $a_0$ and $\kappa$
by a negative result for different impact parameters $b$ (labelling the
lines and coding the colours). For a given $b$, the admissible parameter
space would be ``outside'' the corresponding $b$ line 
(i.e. towards the right lower corner).
}}\end{center}
\end{figure}

%Specifically we may ask, how far would the parameters have to be from their
%fiducial values before a detection becomes hopeless. We address this issue
%by zooming into the problematic region of parameter space (high $\kappa$,
%low $a_0$), plotting the value of the impact parameter $b$ beyond which 
%SNR drops below 1. This is done in Fig.~\ref{fig:null}.

Supposing we get a negative result, what constraints can we place upon 
$a_0$ and $\kappa$? As in~\cite{ali} we may get a preliminary estimate
by seeking the region where the SNR for an optimal filter drops below 1.
This was plotted Fig.~\ref{fig:null} for various values of 
$b$ (in this figure, $b$ labels the lines and codes the colours). 
For a given $b$, the 
admissible parameter space is ``outside'' the corresponding $b$ line 
(i.e. towards the right-bottom corner). In general, a negative result
forces $a_0$ to be smaller and $\kappa$ to be larger than the fiducial
values, the more so, the smaller the impact parameter $b$. As we see,
if we were to miss the saddle by 1500~km or more, the fiducial values
of $a_0$ and $\kappa$ would survive a negative result. For
an approach any closer, however, a negative result
would rule them out and squeeze the parameter space towards the
right-bottom corner. For $b\sim 10\unit{km}$, the $a_0$ (the $\kappa$) 
would have to be smaller (larger) than the fiducial values by an order 
of magnitude. 

These constraints may now be combined with other pressures
upon the theory, such as those arising from limits on $G$ renormalisation,
Big Bang Nucleosynthesis, fifth force solar system tests, 
galaxy rotation curve data, and cosmological structure formation.
However, as advocated in the introduction, by allowing complete freedom
in $a_0$ and $\kappa$ in a saddle test, we have achieved a clear separation
of the issues confronting these theories.

\section{The moon saddle as a LPF target}\label{moon}
Our technique can also be applied to a very topical issue: whether the Moon
saddle is a good alternative target for LPF. Practical matters may render
this saddle more amenable to multiple flybys, an issue that could be essential
in dismissing a ``false alarm'', should a positive detection be found. 
In the absence of a more detailed study of transfer orbits we evaluate 
SNR's for the moon saddle, hoping that this may motivate further work in
orbit design. 

\begin{figure}\begin{center}
\resizebox{1.\columnwidth}{!}{\includegraphics{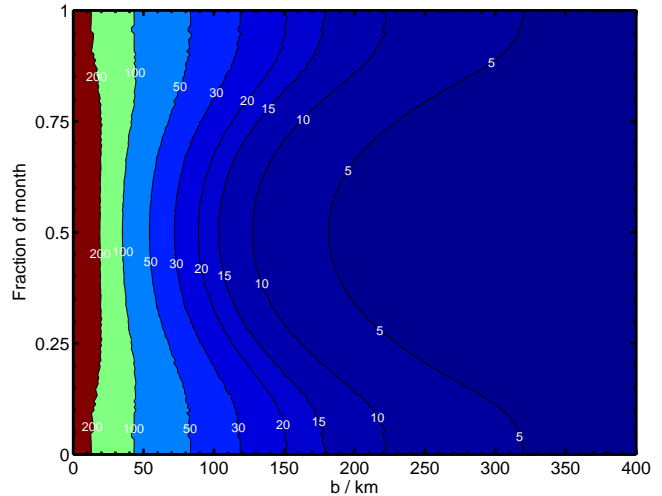}}
\caption{\label{fig:moon}{ SNRs for the moon saddle, 
assuming a standard noise model and speed 
$v=0.3\unit{km}\unit {s}^{-1}$, for different impact parameters 
and day of the month (0 and 1 represent the New Moon,
0.5 the Full Moon). We see that the moon saddle 
is less forgiving if you miss it by more than 150 km and
more rewarding if you get close to it (with SNRs of 200 within reach).
If the former, new moons generate higher SNRs.}}\end{center}
\end{figure}

Application of the algorithm in Section~\ref{scale} to the moon saddle
is straightforward (and indeed it motivated the argument presented therein).
As noted in~\cite{bevis}, $r_0$ for the Moon saddle is smaller than the
380km found for the Earth-Sun saddle, and 
this size is more variable, depending strongly on the phase of the Moon 
(it varies between 25km and 80km; see Fig.10 of~\cite{bevis}).
However $A$ is larger, too, so the tidal stresses have a larger amplitude.
Nevertheless, what really matters for SNRs is the Fourier transform
of the signal as seen in time, with the satellite going through the bubble.
The large SNRs obtained for the Sun-Earth saddle result from a miraculous
coincidence between the sweet spot in the amplitude spectral density (ASD),
and the size of the bubble as transformed into a time-signal by the 
typical velocities found in transfer orbits. This miracle could be spoiled
by the smaller size of the Moon saddle.

As it happens, orbits crossing the Moon saddle do so with a smaller 
velocity, typically smaller than $0.5\unit{km}\unit {s}^{-1}$\footnote{Steve
Kemble, private communication}. The two 
effects---smaller bubble, smaller speed---counteract each other 
when converting the bubble signal into a time signal. Therefore it is
not surprising that the SNRs predicted for the Moon saddle are as high
as those for the Earth saddle, albeit more variable in time, depending
on the phase of the moon. 

In Fig.~\ref{fig:moon} we plotted SNRs assuming the standard noise model
we have used throughout this paper, for a crossing of the moon saddle
at $v=0.3\unit{km}\unit {s}^{-1}$, for different impact parameters 
and day of the month. On the $y$- axis 0 and 1 represent the New Moon,
and 0.5 represents 
the Full Moon. As we can see, in comparison with the Earth-Sun saddle,
the moon saddle:
\begin{itemize}
\item is less forgiving if you miss it by more than 150 km.
\item is more rewarding if you get close to it (with SNRs of 200 within reach).
\item in the first case, then the lunar phase is crucial, with the new moon 
producing the best results.
\end{itemize}
In view of these results we think we should urge the orbit 
designers to include the moon saddle in their considerations.

\section{Conclusions}
To conclude, we have presented a simple argument allowing the inference
of a large database of templates for the tidal stresses that would be felt
by LISA Pathfinder, should a saddle flyby be incorporated into the mission.
The argument allows for the variation of the acceleration scale $a_0$ and
$\kappa$. Should the functional form of the free function 
$\mu$ be changed, the SNRs obtained would change, but only as predicted 
in~\cite{ali}: they wouldn't change much for MONDian 
two-regime functions, unless $b$ is much larger than $r_0$. We may detach these
theories altogether from their ``alternative to dark matter duties''.
Then we may consider two-regime 
functions with $\mu\rightarrow 1$ at large arguments, but
$\mu\propto z^n$, with $n\neq 1$, when $z$ is small. The
scaling argument presented here would still be applicable
in this context, but the fiducial templates
would have to be obtained by re-running a numerical 
code for each $n$. In a paper in preparation we show how this may be
bypassed too, albeit with a much more complex analytical argument.

%Nonetheless a significant amount of labour has been shortcut by
%employing the scaling-law Eqn.~(\ref{ts}), which applies to any $\mu$. 

The practical applications of our technique are far-reaching
and will be the support of a number of future publications concerned
with the data analysis of a saddle test. In this paper we merely showed
how SNRs change by changing the parameters  of the theory. This gives
an indication of how sensitive to them the experiment is, and therefore
how much it will constrain them. More importantly, as an application
we applied our scaling algorithm to the prediction of results for the
Moon saddle. The results were very encouraging and lead us to urge
the orbit designers to include it in their considerations. 

\begin{acknowledgments} We thank Pedro Ferreira and Steve Kemble
for useful discussions as well as  the whole  LPF science team.
AM is funded by an STFC studentship.  All the numerical work was carried out on the COSMOS supercomputer, which is supported by STFC, HEFCE and SGI.
\end{acknowledgments}

\bibliography{references}

\end{document}